\documentclass[11pt]{article}
\usepackage{a4p,cite}
\newcommand{\rb}{\mbox{$R_{\rm b}$}}

\newcommand{\zb}{\mbox{$\rm Z^0$}}

\newcommand{\bbbar}{\mbox{$\rm b\overline{b}$}}

\newcommand{\bqbar}{\mbox{$\rm\overline{b}$}}

\newcommand{\vcb}{\mbox{$V_{\rm cb}$}}
\newcommand{\mvcb}{\mbox{$|V_{\rm cb}|$}}
\newcommand{\mvub}{\mbox{$|V_{\rm ub}|$}}
\newcommand{\vtd}{\mbox{$V_{\rm td}$}}

\newcommand{\mean}[1]{\langle{#1}\rangle}
\newcommand{\meanxe}{\mbox{$\mean{x_E}$}}
\newcommand{\bratio}[2]{\mbox{${\rm Br}(#1\rightarrow #2)$}}
\newcommand{\delm}{\mbox{$\Delta M$}}
\newcommand{\dmd}{\mbox{$\Delta m_{\rm d}$}}
\newcommand{\dms}{\mbox{$\Delta m_{\rm s}$}}

\newcommand{\bplus}{\mbox{$\rm B^+$}}
\newcommand{\bminus}{\mbox{$\rm B^-$}}
\newcommand{\bzero}{\mbox{$\rm B^0$}}
\newcommand{\bzerobar}{\mbox{$\rm\bar{B}^0$}}
\newcommand{\bs}{\mbox{$\rm B_s$}}
\newcommand{\bsbar}{\mbox{$\rm \bar{B}_s$}}

\newcommand{\dstar}{\mbox{$\rm D^{*+}$}}
\newcommand{\ddstar}{\mbox{$\rm D^{**}$}}
\newcommand{\dzero}{\mbox{$\rm D^0$}}
\newcommand{\bztodslv}{\mbox{$\bzerobar\rightarrow\dstar\ell^-\bar{\nu}$}}

\newcommand{\btodshlv}{$\rm \bar{B}\rightarrow\dstar h\,\ell^-\bar{\nu}$}
\newcommand{\btodsslv}{$\rm \bar{B}\rightarrow\ddstar\ell^-\bar{\nu}$}

\newcommand{\taubz}{\mbox{$\rm\tau_{B^0}$}}
\newcommand{\taubp}{\mbox{$\rm\tau_{B^+}$}}
\newcommand{\taubs}{\mbox{$\rm\tau_{B_s}$}}
\newcommand{\ebzero}{\mbox{$E_{\rm B^0}$}}

\newcommand{\pvdstar}{\mbox{$\bf p_{\rm D^*}$}}
\newcommand{\pvdzero}{\mbox{$\bf p_{\rm D^0}$}}

\newcommand{\qopp}{\mbox{$Q_{\rm opp}$}}
\newcommand{\qsame}{\mbox{$Q_{\rm same}$}}
\newcommand{\qvtx}{\mbox{$Q_{\rm vtx}$}}

\newcommand{\qt}{\mbox{$Q_{\rm T}$}}
\newcommand{\qm}{\mbox{$Q_{\rm M}$}}
\newcommand{\ptagu}{\mbox{$P_{\rm u}$}}
\newcommand{\ptagm}{\mbox{$P_{\rm m}$}}
\newcommand{\PLB}[3] {Phys.~Lett.\ {B#1} (#2) #3}
\newcommand{\PRL}[3] {Phys.~Rev.\ {Lett.~#1} (#2) #3}
\newcommand{\PRD}[3] {Phys.~Rev.\ {D#1} (#2) #3}
\newcommand{\PRP}[3] {Phys.~Rep.\ {#1} (#2) #3}
\newcommand{\NIM}[3] {Nucl.~Instrum.\ {Methods~#1} (#2) #3}

\newcommand{\CPC}[3] {Comp.~Phys.\ {Comm.~#1} (#2) #3}
\newcommand{\ZPC}[3] {Z.~Phys.\ {C#1} (#2) #3}
\newcommand{\JPH}[3] {J.~Phys.\ {#1} (#2) #3}
\newcommand{\EPJ}[3] {Eur.~Phys.\ J.\ {C#1} (#2) #3}
\newcommand{\IJA}[3] {Int.~J.~Mod.~Phys.\ A.\ {#1} (#2) #3}

\newcommand{\etal} {et~al.}
\input epsf
\newcommand{\epostfig}[3]{
\begin{figure}[tbp]
\setlength{\epsfxsize}{1.1\hsize}
\hspace*{-0.05\hsize} \epsfbox{#1}
\caption{\label{#2}#3}
\end{figure}
}
\newcommand{\tbzval}{1.541}
\newcommand{\tbzstat}{0.028}
\newcommand{\tbzsyst}{0.023}
\newcommand{\dmdval}{0.497}
\newcommand{\dmdstat}{0.024}
\newcommand{\dmdsyst}{0.025}
\newcommand{\tbdmcorl}{-0.14}
\newcommand{\tbzoval}{1.538}
\newcommand{\tbzostat}{0.025}
\newcommand{\tbzosyst}{0.023}
\newcommand{\dmdoval}{0.479}
\newcommand{\dmdostat}{0.018}
\newcommand{\dmdosyst}{0.015}
\begin{document}
\begin{titlepage}
{\center\Large

EUROPEAN ORGANIZATION FOR NUCLEAR RESEARCH \\

}
\bigskip

{\flushright
% OPAL PR\,317 \\
CERN-EP-2000-90 \\
July 7, 2000 \\
}
\begin{center}
    \LARGE\bf\boldmath
    Measurement of the \bzero\ Lifetime and Oscillation \\
    Frequency using \bztodslv\ decays
\end{center}
\vspace{8mm}
\bigskip

\begin{center}
\Large The OPAL Collaboration \\

\end{center}

\vspace{15mm}

\begin{abstract}
The lifetime and oscillation frequency of the \bzero\ meson has been
measured using \bztodslv\ decays recorded on the \zb\ peak
with the OPAL detector
at LEP. The $\dstar\rightarrow\dzero\pi^+$ decays were reconstructed 
using an inclusive technique and the production flavour of the \bzero\ mesons
was determined using a combination of tags from the rest of the event.
The results
\begin{eqnarray*}
\taubz & = & \tbzval \pm \tbzstat \pm \tbzsyst\,{\rm ps}\, , \\
\dmd & = & \dmdval \pm \dmdstat \pm \dmdsyst\,{\rm ps}^{-1} 
\end{eqnarray*}
were obtained, where in each case the first error is statistical and the
second systematic.
\end{abstract}

\vspace{1cm}     

\begin{center}
\large

Submitted to Physics Letters B.

\end{center}

\end{titlepage}

\begin{center}{\Large        The OPAL Collaboration
}\end{center}\bigskip
\begin{center}{
%begin authorlist PLEASE DO NOT DELETE THIS COMMENT
G.\thinspace Abbiendi$^{  2}$,
K.\thinspace Ackerstaff$^{  8}$,
C.\thinspace Ainsley$^{  5}$,
P.F.\thinspace {\AA}kesson$^{  3}$,
G.\thinspace Alexander$^{ 22}$,
J.\thinspace Allison$^{ 16}$,
K.J.\thinspace Anderson$^{  9}$,
S.\thinspace Arcelli$^{ 17}$,
S.\thinspace Asai$^{ 23}$,
S.F.\thinspace Ashby$^{  1}$
% ...
%end authorlist PLEASE DO NOT DELETE THIS COMMENT
}\end{center}
\newpage
%begin institutes
\noindent $^{  1}$School of Physics and Astronomy, University of Birmingham,
Birmingham B15 2TT, UK
\newline
$^{  2}$Dipartimento di Fisica dell' Universit\`a di Bologna and INFN,
I-40126 Bologna, Italy
\newline
$^{  3}$Physikalisches Institut, Universit\"at Bonn,
D-53115 Bonn, Germany
\newline
% ...
$^{  5}$Cavendish Laboratory, Cambridge CB3 0HE, UK
\newline
% ...
$^{  8}$CERN, European Organisation for Nuclear Research,
CH-1211 Geneva 23, Switzerland
\newline
$^{  9}$Enrico Fermi Institute and Department of Physics,
University of Chicago, Chicago IL 60637, USA
\newline
% ...
$^{ 16}$Department of Physics, Schuster Laboratory, The University,
Manchester M13 9PL, UK
\newline
$^{ 17}$Department of Physics, University of Maryland,
College Park, MD 20742, USA
\newline
% ...
$^{ 22}$Department of Physics and Astronomy, Tel Aviv University,
Tel Aviv 69978, Israel
\newline
$^{ 23}$International Centre for Elementary Particle Physics and
Department of Physics, University of Tokyo, Tokyo 113-0033, and
Kobe University, Kobe 657-8501, Japan
\newline
% ...
%end notes
%
\section{Introduction}

The lifetimes of b hadrons depend both on the strength of the b quark coupling
to the lighter c and u quarks, and on the dynamics of b hadron decay. 
The spectator model prediction that the lifetimes of all heavy hadrons 
containing the same heavy quark are equal is modified by non-spectator
effects dependent on the flavour of the light quark(s) in the hadron.
In contrast to the charm hadrons, where $\tau_{D^+}\approx 2.5\tau_{D^0}$
\cite{pdg98}, 
non-spectator processes are expected to lead to lifetime differences between
the \bplus\ and \bzero\ mesons of at most 10\,\% \cite{btheo}.
Measurements of b hadron
lifetimes at the level  of a few percent or better are therefore needed
to test these predictions, and probe the non-spectator processes contributing
to the decays. In addition, precise measurements of the \bzero\ lifetime
are also needed for the determination of the magnitude of the
CKM matrix element \vcb\ \cite{vcbrev}.

The most precise measurements of the \bplus\ lifetime come from topological
vertex reconstruction techniques, where the selection of charged secondary
vertices allows a clean sample of \bplus\ decays to be isolated
\cite{opalbpbz,topblife}. This method is however limited for \bzero\ lifetime 
measurements, due to 
contamination from other neutral b hadrons (\bs\ mesons and b baryons).
An alternative technique is to use \bztodslv\ decays\footnote{Charge conjugate
reactions are implied, and the symbol $\ell$ refers to either an electron
or muon.}, which can be 
efficiently partially reconstructed by exploiting the low energy release
in the decay $\dstar\rightarrow\dzero\pi^+$. In this case, only the $\pi^+$
from the \dstar\ decay is identified, and no attempt is made to fully
reconstruct the \dzero\ meson decay. This method has previously been 
used by DELPHI \cite{delb0life}
to measure the \bzero\ lifetime, and by DELPHI \cite{delvcb} and
OPAL \cite{opalvcb} to measure \mvcb.

The same sample of \bztodslv\ decays can also be used to measure the 
\bzero\--\bzerobar\ oscillation frequency\footnote{The conventions 
$c=1$ and $\hbar=1$ are employed throughout.} \dmd\ \cite{deldmd}.
In the neutral b meson system, the weak eigenstates \bzero\ and \bzerobar\ 
differ from the mass eigenstates, and transitions between them are possible,
arising dominantly in the Standard Model from second order 
weak transition box diagrams 
involving virtual top quarks. Therefore an initial \bzero\ meson can oscillate
into a \bzerobar\ at time $t$ with a probability given by
\[
P(\bzero\rightarrow\bzerobar)=\frac{1}{2}(1-\cos\dmd t) \, .
\]
Measurements of \dmd\ allow the extraction of the magnitude of the CKM 
matrix element \vtd, though the precision is currently severely limited by 
theoretical uncertainties \cite{dmdth}.

In this paper, measurements of both \taubz\ and \dmd\ based on this
technique are presented. The reconstruction of \bztodslv\ decays
is described in Section~\ref{s:rec}, followed by the determination of the 
proper decay time for each event in Section~\ref{s:propt}, and the 
production flavour tagging needed for the oscillation measurement in 
Section~\ref{s:prodf}. The fit to determine
\taubz\ and \dmd\ is described in Section~\ref{s:fit}, followed by a 
discussion of systematic uncertainties in Section~\ref{s:syst}.
The results are summarised and
combined with previous OPAL measurements in Section~\ref{s:conc}.

\section{Inclusive reconstruction of \bztodslv\ events}\label{s:rec}

The OPAL detector is well described elsewhere \cite{opaldet}.
The data sample used in this analysis consists of about 4 million hadronic 
\zb\ decays collected during the period 1991--1995,
together with an additional 400\,000
events recorded primarily for detector calibration purposes in 1996--2000.
Corresponding simulated event samples
were generated using JETSET 7.4 \cite{jetset} as described in \cite{opalrb}.

Hadronic \zb\ decays were selected using standard criteria 
\cite{opalrb}. To ensure the event was well contained within the 
acceptance of the detector, the thrust axis 
direction\footnote{A right handed coordinate system is used, with positive $z$ 
along the electron beam direction and $x$ pointing to the centre of the LEP
ring. The polar and azimuthal angles are denoted by $\theta$ and $\phi$.} 
was required to satisfy $|\cos\theta_T|<0.9$.
Tracks and electromagnetic calorimeter clusters with no
associated tracks were then combined into jets using a cone
algorithm~\cite{jetcone}, with a cone half-angle of 0.65\,rad and
a minimum jet energy of 5\,GeV. The transverse momentum $p_t$ of each
track was defined relative to the axis of the jet containing it, where
the jet axis was calculated including the momentum of the track.

The reconstruction of \bztodslv\ events was performed by combining
high $p$ and $p_t$ lepton (electron or muon)
candidates with oppositely charged pions
from the $\dstar\rightarrow\dzero\pi^+$ decay. The selection is similar
to that used in \cite{opalvcb}, but with some changes to produce an
unbiased \bzero\ decay proper time measurement and to increase the 
efficiency at the expense of higher combinatorial background.
Electrons were identified and photon conversions rejected using neural
network algorithms \cite{opalrb}, and muons were identified as in 
\cite{muonid}. Both electrons and muons were required to have 
momenta $p>2\rm\,GeV$,
transverse momenta with respect to the jet axis $p_t>0.7\rm\,GeV$, and
to lie in the polar angle region $|\cos\theta|<0.9$. 

The selection of pions from \dstar\ decays relies on the small mass
difference of only 145\,MeV  \cite{pdg98} between the \dstar\ and \dzero, which
means the pions have very little transverse momentum with respect to the
\dzero\ direction. In each jet containing a lepton candidate, 
the \dzero\ momentum vector \pvdzero\ and energy $E_{\rm D^0}$ were
reconstructed using an inclusive technique, selecting
a group of tracks and calorimeter clusters compatible with a \dzero\ decay
using kinematic, impact parameter and invariant mass 
information. This procedure is described fully in \cite{opalvcb} and gives
angular resolutions on the \dzero\ direction of about 45\,mrad in both
$\theta$ and $\phi$.

Each track in the jet (other than the lepton) was then considered in turn
as a slow pion candidate, provided it satisfied $0.5\,{\rm GeV}<p<2.5$\,GeV
and had a transverse momentum with respect to the \dzero\ direction
of less than 0.3\,GeV. If the pion under consideration
had been included in the reconstructed
\dzero, it was removed and the \dzero\ momentum and energy recalculated.
The final selection was made using the reconstructed mass 
difference\footnote{The \dstar--\dzero\ mass difference \delm\
was denoted by $\Delta m$ in \cite{opalvcb}.}
\delm\ between the \dstar\ and \dzero\ mesons, calculated as
\[
\delm=\sqrt{E^2_{\rm D^{*}}-|\pvdstar|^2}-m_{\rm D^0}\, ,
\]
where the \dstar\ energy is given by $E_{\rm D^{*}}=E_{\rm D^0}+E_\pi$
and momentum by $\pvdstar={\bf p}_{\rm D^0}+{\bf p}_\pi$.

The position of the \bzero\ candidate decay vertex was reconstructed
from the intersection point of the lepton and slow pion tracks in the
$x$-$y$ plane. The two-dimensional flight distance of the \bzero\ was 
then calculated as the length of the vector between the $\rm e^+e^-$ 
interaction point (`beamspot')
and the \bzero\ decay vertex, constrained to lie along
the $x$-$y$ projection of the jet direction. This was converted to a 
three dimensional decay distance $L$ using the jet polar angle.
Using just the lepton and pion tracks, together with the interaction point
position and uncertainty determined with a fit to many consecutive events
\cite{beamspot}, results in a decay length estimate which is bias free
and whose resolution does not depend strongly on the decay length itself.
Although the decay length resolution could be improved by adding more tracks
to the \bzero\ decay vertex, this would introduce significant bias at
small decay lengths, and is not necessary as 
the resolution is already adequate for the measurement of \taubz\ and \dmd.

The reconstructed decay length $L$ was signed positive if the \bzero\ decay
vertex was displaced from the beamspot in the direction of the jet momentum,
and negative otherwise. The decay length error $\sigma_L$ was calculated
from the  track parameter and beamspot position error matrices.
The decay length and error were required to satisfy 
$-0.5\,{\rm cm} < L < 2\,{\rm cm}$, $L/\sigma_L>-3$ and $\sigma_L<0.2\rm\,cm$.

The resulting distributions of \delm\ for opposite and same sign lepton-pion
combinations are shown in Figure~\ref{f:dmdmc}(a) and (b).
The predictions of the Monte
Carlo simulation are also shown, broken down into contributions from
signal \bztodslv\ events, `resonant' background containing real leptons 
combined with slow pions from \dstar\ decays, and combinatorial background,  
made up of events with fake slow pions, fake leptons or both.

\epostfig{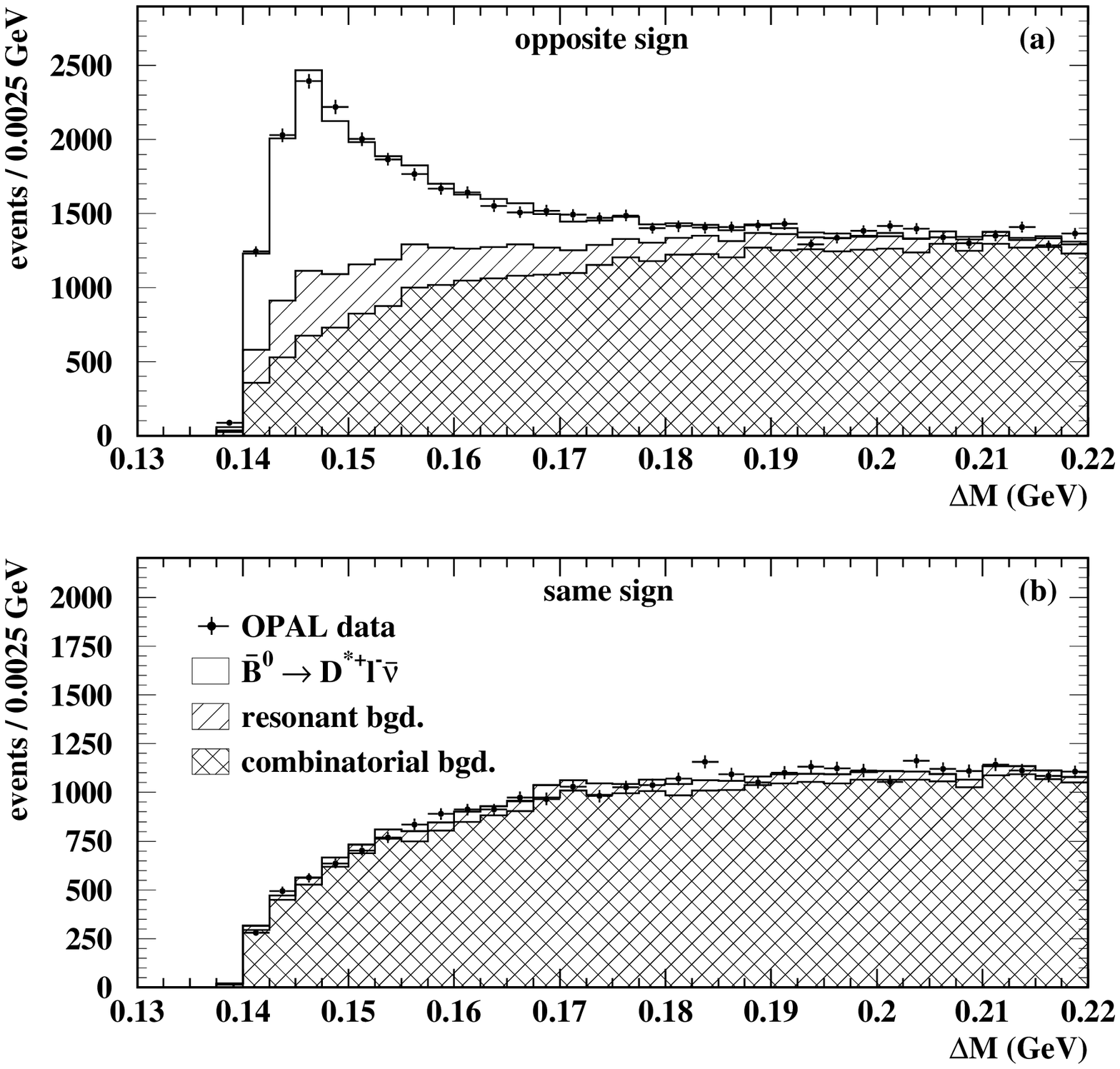}{f:dmdmc}{Reconstructed \delm\ distributions for selected 
(a) opposite sign and (b) same sign lepton-pion combinations. The data are
shown by the points with error bars, and the Monte Carlo simulation
contributions from signal \bztodslv\ decays, other resonant \dstar\ decays 
and combinatorial background are shown by the open, single and cross hatched
histograms respectively.}

In Monte Carlo simulation, about 36\,\% of  opposite sign events with 
$\delm<0.17\rm\,GeV$ are signal 
\bztodslv\ events, 15\,\% are resonant background and
49\,\% are combinatorial background. The 
resonant background is made up mainly of 
$\rm B^-\rightarrow\dstar\pi^-\ell^-\bar\nu$,
$\bzerobar\rightarrow\dstar\pi^0\ell^-\bar\nu$ and
$\bsbar\rightarrow\dstar\rm K^0\ell^-\bar\nu$ decays. These are expected
to be dominated by b semileptonic decays involving orbitally
excited charm mesons (generically referred to as \ddstar),  {\em e.g.\/} 
$\rm B^-\rightarrow D^{**0}\ell^-\bar\nu$ followed by 
$\rm D^{**0}\rightarrow\dstar\pi^-$. These decays will be denoted collectively
by \btodshlv.
Small contributions are also expected
from $\rm b\rightarrow\dstar\tau\bar\nu X$ decays (via any b hadron)
with the $\tau$ decaying leptonically, and 
$\rm b\rightarrow \dstar\rm D_s^-X$ with the $\rm D_s^-$
decaying semileptonically (each about 1\,\% of opposite sign events).
For same sign events with $\delm<0.17\,$GeV, there
is a small resonant contribution of about 6\,\% from events with a real 
$\dstar\rightarrow\dzero\pi^+$ where the \dzero\ decays semileptonically,
and the rest is combinatorial background.

\section{Proper time reconstruction}\label{s:propt}

The proper decay time $t$ of each \bzero\ candidate was calculated 
from its reconstructed decay length $L$ and energy \ebzero. 
The energy was calculated using a technique similar to that described in
\cite{opaldil}, exploiting overall energy and momentum conservation in the
event to account for the missing energy of the unreconstructed neutrino.
The event was treated as a two-body decay of a \zb\ into a \bzero\ of
mass 5.279\,GeV \cite{pdg98} and another object making up the rest of the
event. The \bzero\ energy was calculated as
\[
\ebzero = \frac{E_{\rm cm}^2+m_{\rm B^0}^2-M_{\rm rec}^2}{2 E_{\rm cm}}
\]
where $E_{\rm cm}$ is the centre-of-mass energy of the event and 
$M_{\rm rec}$ the invariant mass of the object recoiling against the \bzero.
The latter was calculated from all tracks and calorimeter clusters in the 
event, excluding
the lepton and those associated to the reconstructed \dstar. A correction
for double counting of charged particles in the tracking detectors
and calorimeters was applied \cite{opalmt}, and the recoil mass
was first scaled by $(\bar E/E_{\rm vis})$ where $E_{\rm vis}$ is
the total event visible energy and $\bar E=87\,\rm GeV$ is the typical 
visible energy in events with only one neutrino. This procedure
improves the resolution in events
where a second neutrino is present \cite{opalbinc}.
The resulting energy estimate is unbiased and has an RMS resolution
of 3.8\,GeV in Monte Carlo \bztodslv\ events, as shown in 
Figure~\ref{f:resfn}(a).

\epostfig{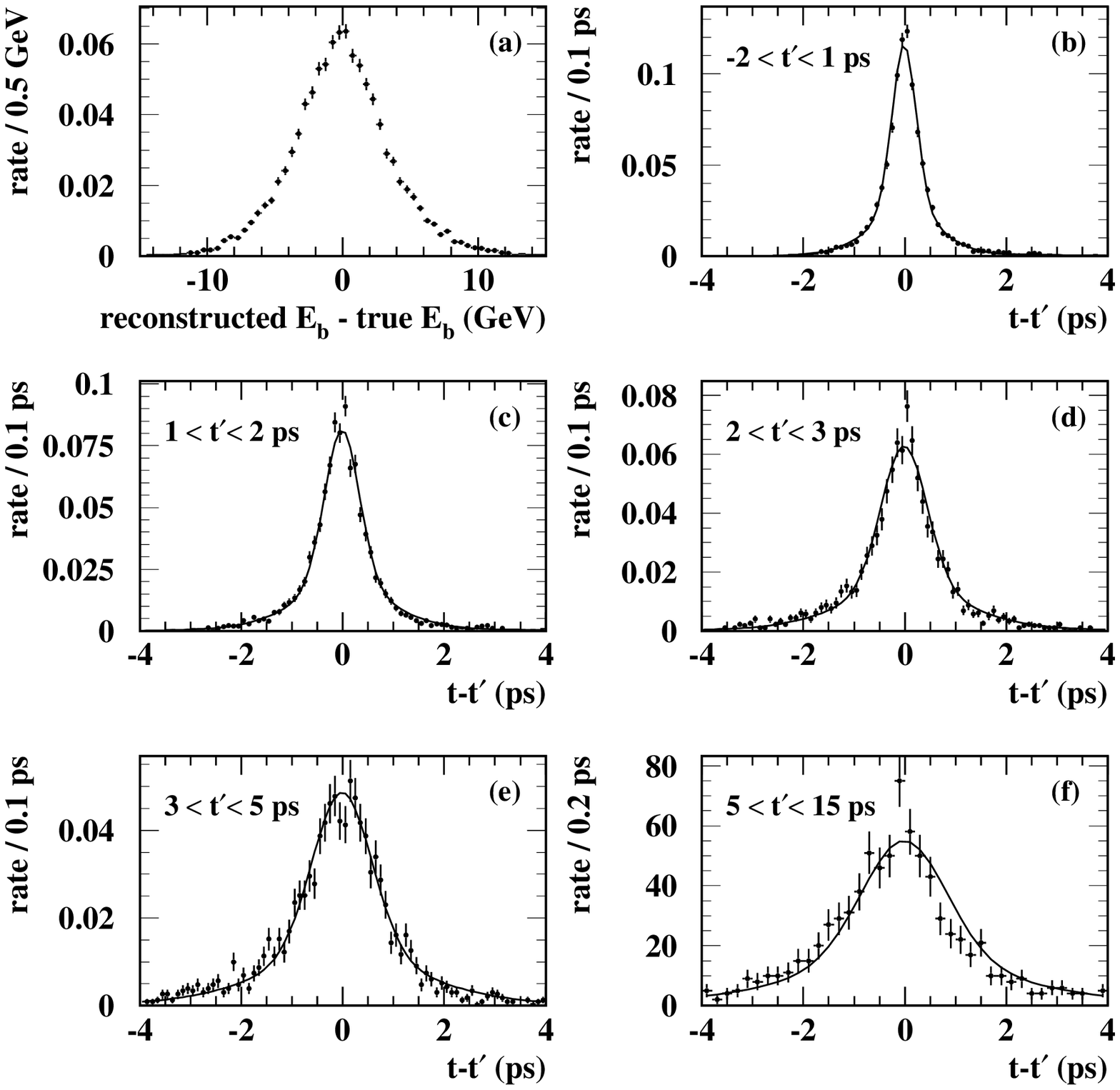}{f:resfn}{(a) Reconstructed b hadron energy resolution
and (b--f) reconstructed proper time 
resolution in various ranges of true proper time $t'$, for Monte Carlo
\bztodslv\ events. The Monte Carlo is shown by the points with error bars,
and the resolution function is shown by the solid line in (b--f).}

The proper decay time $t$ was then calculated from the candidate decay length
and energy as
\[
t=\frac{m_{\rm B^0}L}{\sqrt{E_{\rm B^0}^2-m_{\rm B^0}^2}}\, .
\]
The resulting proper time resolution depends on the true proper
time $t'$, but has no significant bias, as shown in Figure~\ref{f:resfn}(b--f).
The resolution degrades with increasing proper time due to the larger
influence of the energy resolution at large decay lengths.
The resolution was parameterised as a function $R_{\rm D^{*+}}(t,t')$ 
giving the expected distribution of reconstructed $t$ for each true value $t'$.
The resolution function was implemented as the sum of three
Gaussian distributions,
whose widths vary linearly with $t'$. The resolution function is also shown
as the solid line in Figure~\ref{f:resfn}(b--f), and gives a reasonable
description of the Monte Carlo resolution, adequate for measuring the lifetime
and relatively slow \bzero\ meson oscillations.

Similar resolution functions $R_{\rm D^{*+}\pi^-}(t,t')$,
$R_{\rm D^{*+}\pi^0}(t,t')$ and $R_{\rm D^{*+}K^0}(t,t')$
were generated for the three main \ddstar\ background contributions 
$\rm B^-\rightarrow\dstar\pi^-\ell^-\bar\nu$,
$\bzerobar\rightarrow\dstar\pi^0\ell^-\bar\nu$ and
$\bsbar\rightarrow\dstar\rm K^0\ell^-\bar\nu$. In these events, the 
b hadron energy reconstruction has a bias of about $-1.5\rm\,GeV$ as some of
the b hadron decay products are incorrectly included in the recoil mass.
This bias was corrected in the corresponding resolution functions.

\section{Production flavour tagging}\label{s:prodf}

The \bzero\ oscillation frequency measurement requires that the production
and decay flavour (\bzero\ or \bzerobar) of each meson be determined, in 
addition to the decay proper time $t$. The decay flavour can be 
determined from 
the sign of the lepton in the \bztodslv\ decay, but the production flavour
must be determined from other information in the event. The production of
quark anti-quark pairs in $\zb\rightarrow\bbbar$ decays allows the
production flavour of the \bzero\ to be inferred from that of the other 
b hadron in the event. This information is diluted due to 
the possible mixing of the second b hadron.
The other particles in the \bzero\ jet (produced in
the fragmentation of the b quark) also carry some useful information.

The tagging information was extracted using the methods described in
\cite{opalbpbz,opaljpks}. The event was divided into two hemispheres by
the plane perpendicular to the thrust axis and containing the $\rm e^+e^-$
interaction point. In the hemisphere opposite to that containing the \bzero\
jet, up to three pieces of information were used to tag the b hadron flavour:
\begin{itemize}
\item The jet charge \qopp\ of the highest energy jet in the opposite
hemisphere, defined as
\begin{equation}\label{e:jetc}
\qopp=\frac{\sum_i (p^l_i)^\kappa q_i}{\sum_i (p^l_i)^\kappa}
\end{equation}
where $p^l_i$ is the longitudinal momentum component with respect to the
jet axis and $q_i$ the charge of track $i$, and the sum is taken over
all tracks in the jet. The parameter $\kappa$ was set to 0.5, as in 
\cite{opaljpks}.

\item The charge \qvtx\ of a secondary vertex (if existing), 
reconstructed in any jet in the opposite hemisphere, as in \cite{opalbpbz}.
A well reconstructed charged vertex indicates a \bplus\ or \bminus\ hadron,
tagging the parent quark as a \bqbar\ or b respectively, whilst a neutral
or badly reconstructed vertex gives no information on the b quark flavour.

\item The charge of a high momentum lepton found in any jet in the opposite 
hemisphere, selected as in \cite{opalbpbz}. A high momentum lepton is
most likely to come from a b hadron decay, again tagging the parent 
b or \bqbar\ quark according to its charge. A neural network algorithm
was used to suppress fake leptons and those coming from cascade charm decays
($b\rightarrow c\rightarrow\ell$) which have the wrong charge correlation.
\end{itemize}

These variables were combined using a neural network algorithm into 
a single tagging variable \qt\ for the opposite hemisphere 
\cite{opalbpbz,opaljpks}, quantifying the confidence with which 
the hemisphere was tagged as containing a b or \bqbar\ hadron. Different
neural networks were used depending on what combination of vertex and/or
lepton variables were available to combine with the jet charge \qopp.

In the hemisphere containing the \bzero\ jet, only the jet charge \qsame\ can
be used to infer the \bzero\ production flavour. \qsame\ was calculated
using equation~\ref{e:jetc}, but with the parameter $\kappa$ set to zero, so
it becomes simply the average of the charges of the tracks in the jet.
This avoids being sensitive to the decay flavour of the \bzero\ (and hence 
whether it has mixed or not), but is still sensitive to the production flavour
via the information carried by the fragmentation tracks in the 
jet \cite{opaljetc}. The jet charge \qsame\ was used to generate a second
hemisphere tagging variable \qm\, independent of the opposite hemisphere
variable \qt.

The two tagging variables were combined to produce a single tag
$Q_2$ for the \bzero\ production flavour, as in \cite{opalbpbz}. 
The continuous variable $Q_2$ ranges from $-1$ to $+1$, and is defined 
such that events with $Q_2=+1$ are tagged 
with complete confidence as containing a produced \bzero, events
with $Q_2=-1$ are tagged with complete confidence as containing a
produced \bzerobar, and events with $Q_2=0$ are equally likely to be 
\bzero\ or \bzerobar. The modulus $|Q_2|$ satisfies $|Q_2|=1-2\eta$, where
$\eta$ is the `mis-tag' probability, {\em i.e.\/} the probability to
tag the production flavour incorrectly.
Finally, the production flavour tag $Q_2$ and \bzero\ decay lepton
sign $l$ were combined to produce the mixing tag $Q=Q_2\cdot l$, such that
events with $Q>0$ ($Q<0$) are tagged as unmixed (mixed).

Both the sign and magnitude of $Q$ are used in the fit to determine
\dmd, giving the events with high probability to be correctly tagged more
weight. Considering only the sign of $Q$, 31\,\% of signal events are tagged
incorrectly. The event-by-event weighting reduces
this to an effective mis-tag of 28\,\%, equivalent to a 33\,\% increase
in statistical sensitivity.

\section{Fit and results}\label{s:fit}

The values of \taubz\ and \dmd\ were extracted using an unbinned
extended maximum likelihood fit to the reconstructed mass difference
\delm, proper time $t$  and mixing tag $Q$ of each event. 
Both opposite and same sign events 
with $\delm<0.22\rm\,GeV$ and $-2<t<15\,$ps
were used in the fit, the high \delm\ and same sign 
events serving to constrain the combinatorial background 
normalisation and shapes in the opposite sign low \delm\ 
region populated by the \bztodslv\ decays. Using the \delm\ value from
each event in the fit, rather than just dividing the data into 
low \delm\ `signal' and high \delm\ 
`sideband' mass regions, increases the statistical sensitivity as the
signal purity varies considerably within the low \delm\ region.

The likelihood is similar to that used in \cite{opalvcb} for the
measurement of \mvcb. The logarithm of the overall likelihood was given by
\begin{equation}\label{e:totlike}
\ln {\cal L} = \sum_{i=1}^{M^a} \ln {\cal L}^a_i +
\sum_{j=1}^{M^b} \ln {\cal L}^b_j -N^a-N^b
%{\cal L} = e^{-N^r-N^w} \left( \prod_{i=1}^{M^r} {\cal L}^r_i \right) 
%\left( \prod_{j=1}^{M^w} {\cal L}^w_j \right)
\end{equation}
where the sums of individual event log-likelihoods $\ln {\cal L}^a_i$ and 
$\ln {\cal L}^b_j$ are taken over all the observed $M^a$ opposite sign and
$M^b$ same sign events in the data sample, and $N^a$ and $N^b$ are the
corresponding expected numbers of events.

The likelihood for each opposite sign event was given in terms of 
different types or sources of event by
\begin{equation}\label{e:levtr}
{\cal L}^a_i(\delm_i,t_i,Q_i) 
= \sum_{s=1}^4 N^a_s  M_s(\delm_i) \, T_s(t_i,Q_i)
\end{equation}
where $N^a_s$ is the number of expected events,
$M_s(\delm)$ the mass difference distribution
and $T_s(t,Q)$ the proper time distribution
for source $s$. For each source, the mass difference distribution
$M_s(\delm)$ is normalised to
one. The total number of expected events is given by the sum of the
individual contributions: $N^a=\sum_{s=1}^4 N^a_s$.

There are four opposite sign sources: (1) signal \bztodslv\ events, (2)
\btodshlv\ events where the \dstar\ is produced via
an intermediate \ddstar, (3) other opposite sign background
involving a genuine lepton and a slow pion from \dstar\ decay  and 
(4) combinatorial background. The sum of sources 2 and 3 are shown as
`resonant background' in Figure~\ref{f:dmdmc}.
A similar expression to equation~\ref{e:levtr} was used for ${\cal L}^b_j$,
the event likelihood for same sign events. In this case, only  sources~3
and~4 contribute.

The mass difference distributions $M_s(\delm)$ 
for sources 1--3 were represented by analytic functions, whose parameters
were determined using large numbers of simulated events.
For the signal (source 1), both unmixed and mixed events must be considered
to determine the proper time distribution $T_1(t,Q)$. The probability
to find a signal decay with true proper time $t'$ is given by:
\[
P_1(t',Q)=\frac{1}{\taubz}\, e^{-t'/\taubz} \left(
\ptagu(Q)\frac{(1+\cos\dmd t')}{2}+\ptagm(Q)\frac{(1-\cos\dmd t')}{2} \right)
\]
where \ptagu\ (\ptagm) is the probability that the event is unmixed (mixed)
given the observed mixing tag $Q$. These probabilities are given by
$\ptagu(Q)=(1+Q)/2$ and $\ptagm(Q)=(1-Q)/2$. The signal probability is then 
convolved with the time resolution function $R_{\rm D^{*+}}(t,t')$ 
described in Section~\ref{s:propt} to 
give the expected reconstructed proper time distribution 
$T_1(t,Q)$ as a function of the assumed \bzero\ lifetime \taubz\ 
 and oscillation frequency \dmd:
\begin{equation}\label{e:tsig}
T_1(t,Q)=\int_0^\infty dt' \frac{1}{2\taubz}\, e^{-t'/\taubz}
(1+Q\cos\dmd t') R_{\rm D^{*+}}(t,t')\, .
\end{equation}
The corresponding proper time distribution $T_2(t,Q)$ for source~2 
(\ddstar) was calculated from the sum of the individual contributions
from \bplus, \bzero\ and \bs\ decays via \ddstar. For \bzero\ and 
\bs\ decays, equation~\ref{e:tsig} was modified, with the resolution
function $R_{\rm D^{*+}}(t,t')$ replaced by $R_{\rm D^{*+}\pi^0}(t,t')$
(\bzero\ decays) or $R_{\rm D^{*+}K^0}(t,t')$ (\bs\ decays), and 
\taubz\ and \dmd\ replaced by \taubs\ and \dms, the \bs\ lifetime and
oscillation frequency, for \bs\ decays. For \bplus\ decays, which are always
unmixed, the proper time distribution takes the form
\[
T_{2,\rm B^+}(t,Q)=\int_0^\infty dt' \frac{1}{2\taubp}\, e^{-t'/\taubp}
(1+Q) R_{\rm D^{*+}\pi^-}(t,t')\, .
\]
The values of \taubp, \taubs\ and \dms\ and their corresponding uncertainties
were taken from \cite{pdg98,pdg99}
and are given in Table~\ref{t:input}; \dms\ was varied between the
experimental lower limit and 50\,ps$^{-1}$, well above the experimental
sensitivity and theoretically favoured region.

The number of signal events $N^a_1$ was left as a free parameter in the fit.
The number of events in source~2, and the relative contributions of 
\bplus, \bzero\ and \bs\ decays within this source, were calculated 
as described in \cite{opalvcb}, from the measured branching ratio
$\bratio{\rm b}{\dstar\pi^-\ell\bar\nu X}=(0.473\pm 0.095)\,\%$ 
\cite{alephdss} and the assumption of isospin and SU(3) symmetry.
This was combined with the selection efficiency estimated from Monte Carlo
and the values of \rb\ and \bratio{\dstar}{\dzero\pi^+} listed in 
Table~\ref{t:input} to determine the expected number of these
events in the data sample.

\begin{table}
\centering

\begin{tabular}{l|c|l}
Quantity & Assumed value & Reference \\ \hline
\taubp\ & $1.65\pm 0.03$\,ps & \cite{pdg99} \\
\taubs\ & $1.54\pm 0.07$\,ps & \cite{pdg98} \\
\dms\ & 9.1--50\,ps$^{-1}$ & \cite{pdg99}, see text \\
\rb\ & $(21.70\pm 0.09)\,\%$ & \cite{pdg98} \\
\bratio{\dstar}{\dzero\pi^+} & $(68.3\pm 1.4)\,\%$ & \cite{pdg98} \\
\bratio{\rm b}{\dstar h\,\ell\bar\nu} & $(0.76\pm 0.16)\,\%$ & 
\cite{alephdss,opalvcb}\\
\bratio{\rm b}{\dstar\tau^-\bar\nu X} & $(0.65\pm 0.13)\,\%$ & 
\cite{pdg98,opalvcb} \\
\bratio{\bzerobar}{\dstar\rm D_s^{(*)-}} & $(4.2\pm 1.5)\,\%$ & \cite{pdg98}\\
\bratio{\rm b}{\dstar X} & $(17.3\pm 2.0)\,\%$ & \cite{opaldprod} \\
\bratio{\rm c}{\dstar X} & $(22.2\pm 2.0)\,\%$ & \cite{opaldprod} \\
\hline
\end{tabular}
\caption{\label{t:input} Input quantities used in the fit for
\taubz\ and \dmd.}
\end{table}

The numbers $N_3^{a,b}$ of events in the small 
background contributions covered by source~3 (both opposite and same sign)
were taken from Monte Carlo simulation, with branching ratios adjusted to the
values given in Table~\ref{t:input}, as described in more detail in 
\cite{opalvcb}. The proper time distributions $T_3(t,Q)$ 
were described by negative exponentials convolved with Gaussian resolution 
functions, with parameters again determined from Monte Carlo, 
as were the fractions of mixed events in each contribution.

The parameters of the analytic functions describing the combinatorial 
background ($N_4^a$, $N_4^b$, $M_4(\delm)$ and
$T_4(t,Q)$ were fitted entirely from the data, with only the choice
of functional forms motivated by simulation. The shapes of the
mass and proper time functions (including a small correlation between
\delm\ and $t$) are constrained by the same sign sample (which is
almost entirely combinatorial background), and the opposite sign high 
\delm\ region serves to normalise the number of combinatorial background 
events in the low \delm\ region. Since the combinatorial background
is dominated by semileptonic b decays, it contains oscillating components
in both opposite and same sign samples. The amplitude and frequency
of this oscillation were constrained to be the same in both samples,
but the fraction of unmixed events was allowed to be different, as supported
by Monte Carlo simulation  studies.

The values of \taubz\ and \dmd\ were extracted by maximising the total
likelihood given by equation~\ref{e:totlike}. The values of
\taubz\ and \dmd\ were allowed to vary, together with the number of signal
events and 18 auxiliary parameters describing the combinatorial background
level, mass difference, time and mixing distributions.
A result of 
\begin{eqnarray*}
\taubz & = & \tbzval \pm \tbzstat\,{\rm ps}\, , \\
\dmd   & = & \dmdval \pm \dmdstat\,{\rm ps^{-1}}
\end{eqnarray*}
was obtained, where the errors are only statistical. The correlation
between \taubz\ and \dmd\ is $\tbdmcorl$. The fitted values of
the other parameters were consistent with expectations from Monte Carlo
simulations.

The distributions
of reconstructed proper time $t$ for opposite and same sign data events with 
$\delm<0.17$\,GeV, together with the fit results, are shown
in Figure~\ref{f:propt}. The fractions of mixed events $R$,
corrected for the dilution due to mis-tagged events, are shown for the
same \delm\ regions in Figure~\ref{f:rplot}, where the oscillation of signal
\bztodslv\ decays and  the oscillation in the combinatorial background
are clearly visible.
The fit describes the data well, both in this mass region 
and in the high \delm\ region dominated by combinatorial background.
The discrepancy at high reconstructed proper times in the opposite sign
sample (Figure~\ref{f:propt}(a)) is caused by imperfections in the resolution
function (see Figure~\ref{f:resfn}(f)) and is addressed as a systematic
error.

\epostfig{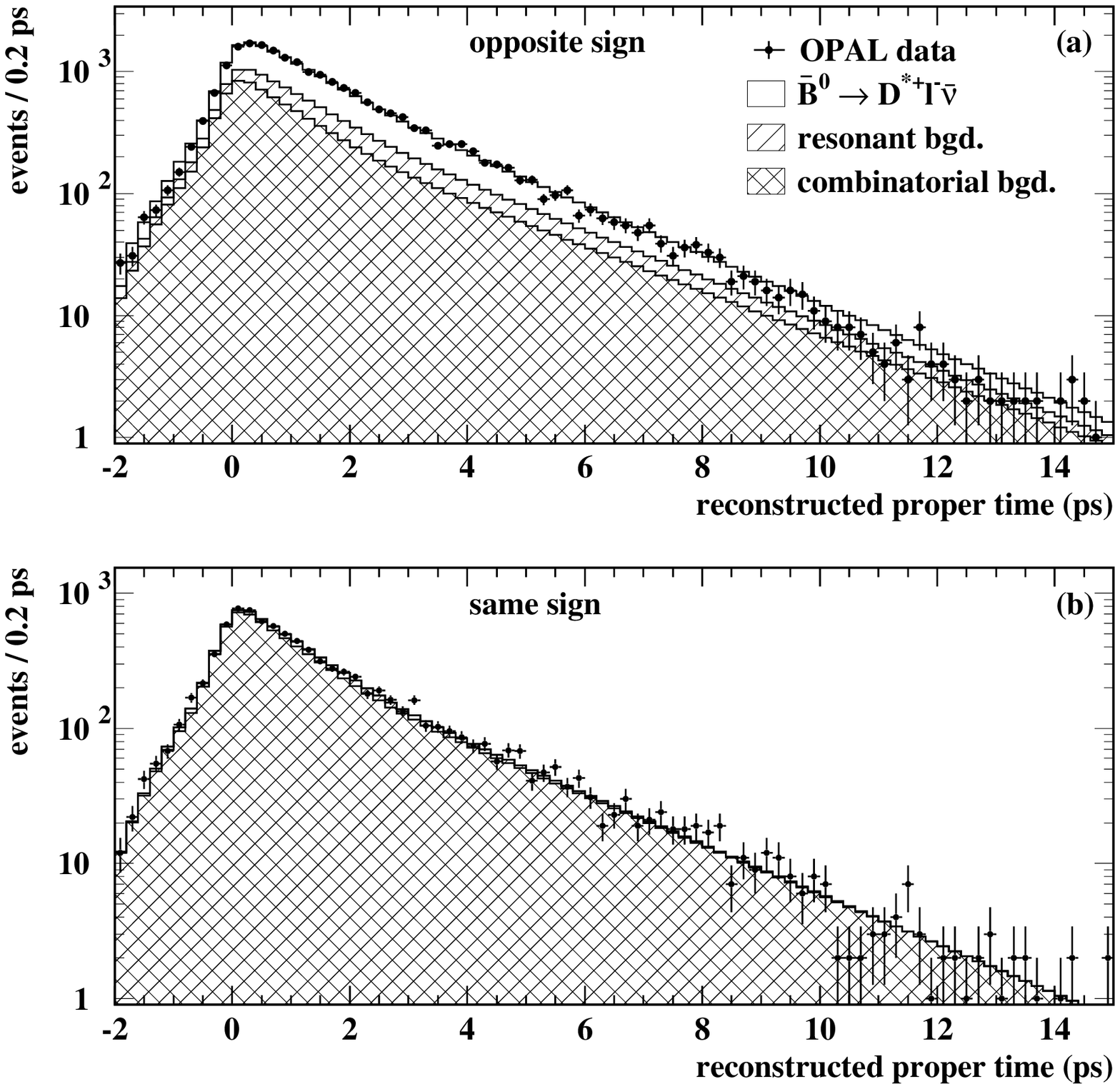}{f:propt}{Distributions of reconstructed proper time 
$t$ for (a) opposite sign and (b) same sign events with $\delm<0.17\,$GeV. The
data are shown by the points with error bars and the expectation from the 
fit result by the histograms. The contributions from signal \bztodslv, resonant
and combinatorial backgrounds are indicated.}

\epostfig{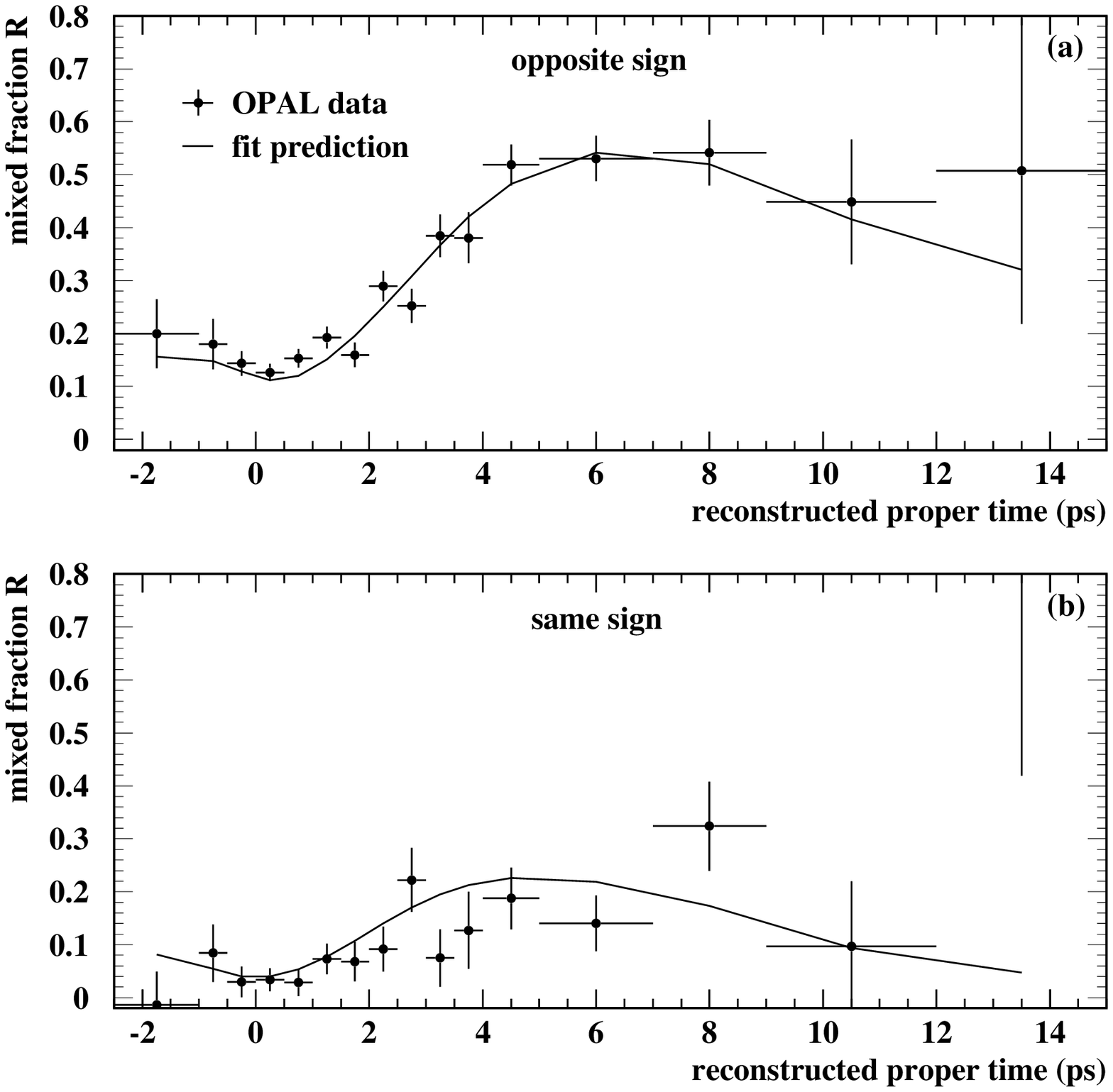}{f:rplot}{Corrected mixed event fractions $R$ {\em vs.\/}
reconstructed proper time $t$ for (a) opposite sign and (b) same sign events
with $\delm<0.17\,$GeV. The data are shown by the points with error bars
and the expectation from the fit result by the solid line.}

\section{Systematic Errors}\label{s:syst}

Systematic errors arise from the uncertainties in the fit input parameters
given in Table~\ref{t:input}, the Monte Carlo modelling of the
signal proper time  resolution and backgrounds, the production flavour
tagging performance and possible biases in the fitting method.
The resulting systematic errors on the values of \taubz\ and \dmd\
are summarised in Table~\ref{t:syst} and described in more detail below.

\begin{table}[tp]
\centering

\begin{tabular}{l|c|c}
Error Source & $\Delta(\taubz)$ (ps) & $\Delta(\dmd)$ (ps$^{-1}$) \\ 
\hline
\bplus\ lifetime                             & 0.004 & 0.001 \\
\bs\ lifetime                                & 0.002 & 0.001 \\
\bratio{\rm b}{\dstar\tau^-\bar{\nu}}        & 0.001 & 0.001 \\
\bratio{\rm b}{\dstar\rm D_s^{(*)-}}         & 0.005 & 0.002 \\
\bratio{\rm b}{\dstar X}                     & 0.001 & 0.001 \\
$\rm b\rightarrow\ddstar\ell\bar{\nu}X$ decays & 0.004 & 0.009 \\
Other backgrounds                            & 0.006 & 0.005 \\
b fragmentation                              & 0.005 & 0.014 \\
\dzero\ decay multiplicity                   & 0.004 & 0.003 \\
Flavour tagging offsets                      & 0.000 & 0.001 \\
Flavour tagging mis-tag                      & 0.000 & 0.009 \\
Tracking resolution                          & 0.015 & 0.010 \\
Detector alignment                           & 0.003 & 0.003  \\
Fit method                                   & 0.012 & 0.010  \\
\hline
Total & \tbzsyst & \dmdsyst 
\end{tabular}
\caption{\label{t:syst}
Summary of systematic errors on the measured values of
\taubz\ and \dmd.}
\end{table}

\begin{description}
\item[Input quantities:] The various numerical fit inputs were each
varied according to the errors given in Table~\ref{t:input} 
and the fit repeated to assess the resulting uncertainties. The \bs\ 
oscillation frequency \dms\ was varied from the experimental
lower limit of $9.1\rm\,ps^{-1}$ \cite{pdg99} to $50\rm\,ps^{-1}$, with
no significant effect on the fit results.

\item[$\bf\boldmath\rm b\rightarrow\ddstar\ell\bar\nu$ decays:] 
The systematic error due to the $\rm b\rightarrow\ddstar\ell\bar\nu$ background
is dominated by uncertainties in the rate of such decays, determined
from the measured $\bratio{\rm b}{\dstar\pi^-\ell\bar\nu X}$ as discussed
in Section~\ref{s:fit}. To evaluate possible
dependence in the $\rm b\rightarrow\ddstar\ell\bar\nu$ decay model, 
the default JETSET model used in the simulation was 
reweighted according to the calculations of Leibovich~\etal \cite{ligeti}
as used in \cite{opalvcb}, with negligible change in the fit results.
The relative contributions of \bplus\ and \bzero\ decays to this sample
are expected to be predicted well by isospin symmetry, but the \bs\ 
contribution was scaled by $0.75\pm 0.25$ relative to the
isospin prediction to account for possible SU(3) flavour violation effects,
as in \cite{opalvcb}. Again, the resulting changes in the fit results
were negligible.

\item[Other backgrounds:] The backgrounds in fit source~3 from 
opposite sign $\rm b\rightarrow\dstar\tau\bar\nu X$ and 
$\rm b\rightarrow\dstar\rm D_s^-X$, and same sign 
$\dstar\rightarrow\dzero\pi^+$,
$\dzero\rightarrow\ell^+X$ decays have characteristic lifetimes of
1.87\,ps and 1.42\,ps respectively in the simulation. These values depend
both on the true b hadron lifetimes and the mis-reconstruction of the
apparent b hadron energy in these events, and were conservatively varied 
by $\pm 0.1\,$ps to assess systematic errors. 
These backgrounds also have oscillating
components due to the contributions from \bzero\ decays (typically 60\,\% of
the events). The systematic error was assessed by varying the \bzero\ fraction
up to 100\,\% of the contribution from b hadrons, and repeating the fit.

\item[b fragmentation:] The effect of uncertainties in the average 
b hadron energy $\meanxe=E_{\rm b}/E_{\rm beam}$ was assessed in Monte Carlo
simulation
by reweighting the events so as to vary \meanxe\ in the range $0.702\pm 0.008$
as recommended by the LEP electroweak working group 
\cite{hfew}, and repeating the fit. This value is also consistent with a 
recent determination of $\meanxe=0.714\pm 0.009$ from SLD \cite{sldbfrag}.
The variation was implemented using the fragmentation 
functions of Peterson, Collins and Spiller,
Kartvelishvili and the Lund group \cite{fragall}, and the largest
observed variations were taken as systematic errors.

\item[$\bf\boldmath\dzero$ decay multiplicities:] 
The b hadron energy reconstruction is sensitive to 
the \bzero\ decay multiplicity, which depends only 
on the \dzero\ decay for the \bzero\ decay channels of interest.
The systematic error was assessed by varying
separately the \dzero\ charged and $\pi^0$ decay multiplicities in 
the Monte Carlo simulation
according to the measurements of Mark III \cite{markIII}. The branching
ratio $\dzero\rightarrow\rm K^0$ or $\rm\bar{K}^0$ 
was also varied according to its uncertainty \cite{pdg98}. 
The resulting uncertainties on \taubz\ and \dmd\ 
from each variation were added in quadrature to determine the total
systematic errors.

\item[Production flavour tagging:]

The jet and vertex charge distributions \qopp, \qsame\ and \qvtx\
are not charge symmetric because of detector effects causing a difference
in the rate and reconstruction of positive and negative tracks 
\cite{opallqjet,opalbpbz}. These offsets were measured directly from the
data as in \cite{opalbpbz}, and their statistical uncertainties contribute
to the systematic error on \dmd.

The production flavour tagging mis-tag estimates must also be 
correct---{\em i.e.\/} that the values of $Q$ correctly represent the mis-tag
probabilities in the data. Excluding the events tagged by jet charge alone,
the opposite flavour tags were checked in \cite{opalbpbz} to a precision
of $\pm 2.6\,\%$, using events tagged in both hemispheres. The same technique
has been used to check the jet charge \qopp, which was found to be 
correct to a precision of $\pm 10\,\%$. The same hemisphere jet charge 
tag \qsame\ was also checked in \cite{opalbpbz} to $\pm 10\,\%$. These
uncertainties were translated into systematic errors on \dmd\ by 
scaling the values of \qt\ and \qm\ and repeating the fit.

\item[Tracking resolution:] Uncertainties in the tracking
detector resolution affect the time resolution for 
\bztodslv\ and \btodsslv\ events. The associated error
was assessed in the simulation by applying a global
10\,\% degradation to the angular and impact parameter resolutions
of all tracks, independently in the $r$-$\phi$ and
$r$-$z$ planes as in \cite{opalrb}, and repeating the fit.
%and removing 1\,\% of $r$-$\phi$ and 
%3\,\% of $r$-$z$ associated silicon micro vertex detector hits to cover 
%residual discrepancies in the Monte Carlo simulation of the silicon hit 
%matching efficiency \cite{opalrb}.

\item[Detector alignment:] The results are sensitive to the 
effective radial positions of the silicon detector wafers (both the positions
of the detectors themselves and the positions of the charge collection
regions within them), which are known to a precision $\pm 20\rm\,\mu m$ 
from studies of cosmic ray events \cite{opalrb}. The resulting uncertainty
was assessed in Monte Carlo simulation by displacing one or both
microvertex detector barrels by $20\rm\,\mu m$ and repeating the fit.

\item[Fit method:] The entire fitting procedure was tested on a fully
simulated Monte Carlo sample six times bigger than the data, with true
values of $\taubz=1.6$\,ps and $\dmd=0.437\rm\,ps^{-1}$. The fit gave the 
results $\taubz=1.604\pm 0.012$\,ps and $\dmd=0.440\pm 0.010\rm\,ps^{-1}$,
in both cases consistent with the true values. The statistical errors on the
fitted values were taken as systematic errors to account for
possible biases in the fit.
Additionally, the large Monte Carlo sample was reweighted to change
the values of \taubz\ and \dmd, and the fit correctly recovered the modified
values. To verify the correctness of the statistical errors returned by the 
fit, the large Monte Carlo sample was split into many subsamples,
and the distribution of fitted results studied.
The signal mass distribution $M_1(\delm)$ was varied to
 cover possible discrepancies in the Monte Carlo modelling of the
signal mass spectrum (see Figure~\ref{f:dmdmc}). 
The data fit was repeated changing the minimum reconstructed 
time cut to $-1.5\,$ps and the maximum to $10\,$ps 
(see Figure~\ref{f:propt}(a)), and  varying the maximum
mass difference \delm\ between 0.20 and 0.25\,GeV. Finally, the data was 
divided up into 5~subsamples according to the year 
of data taking. In all cases, consistent results were found, and no additional
systematic error was assigned.
\end{description}

\section{Conclusions}\label{s:conc}

The lifetime and oscillation frequency of the \bzero\ meson have been
measured using \bztodslv\ decays selected with an inclusive
technique. The results are:
\begin{eqnarray*}
\taubz & = & \tbzval \pm \tbzstat \pm \tbzsyst\,{\rm ps}\, , \\
\dmd & = & \dmdval \pm \dmdstat \pm \dmdsyst\,{\rm ps}^{-1} 
\end{eqnarray*}
where in each case the first error is statistical and the
second systematic. These results are consistent with other determinations
of \taubz\ \cite{opalbpbz,topblife,delb0life,opaltbex,othertb0}
and \dmd\ \cite{deldmd,opaldil,opallqjet,opalexdmd,otherdmd},
with the world averages of
$\taubz=1.54\pm 0.03$\,ps and $\dmd=0.470\pm 0.019$\,ps$^{-1}$ \cite{pdg99}, 
and are in each case the most precise determinations to date.

The results presented here have been combined with the previous 
OPAL measurements of \taubz\ using neutral secondary vertices \cite{opalbpbz}
and exclusively reconstructed $\bzero\rightarrow\rm D^{(*)}\ell\bar\nu$ 
decays \cite{opaltbex} and with measurements
of \dmd\ from single lepton \cite{opallqjet},  dilepton \cite{opaldil}
and exclusively reconstructed \dstar\ decays \cite{opalexdmd}, taking into
account statistical and systematic correlations.
% using the methods described in \cite{comb}. 
The results
\begin{eqnarray*}
\taubz & = & \tbzoval \pm \tbzostat \pm \tbzosyst\,{\rm ps}\, , \\
\dmd & = & \dmdoval \pm \dmdostat \pm \dmdosyst\,{\rm ps}^{-1} 
\end{eqnarray*}
were obtained, where again the first errors are statistical and the
second systematic.

\section*{Acknowledgements}

We particularly wish to thank the SL Division for the efficient operation
of the LEP accelerator at all energies
 and for their continuing close cooperation with
our experimental group.  We thank our colleagues from CEA, DAPNIA/SPP,
CE-Saclay for their efforts over the years on the time-of-flight and trigger
systems which we continue to use.  In addition to the support staff at our own
institutions we are pleased to acknowledge the  \\
Department of Energy, USA, \\
National Science Foundation, USA, \\
Particle Physics and Astronomy Research Council, UK, \\
Natural Sciences and Engineering Research Council, Canada, \\
Israel Science Foundation, administered by the Israel
Academy of Science and Humanities, \\
Minerva Gesellschaft, \\
Benoziyo Center for High Energy Physics,\\
Japanese Ministry of Education, Science and Culture (the
Monbusho) and a grant under the Monbusho International
Science Research Program,\\
Japanese Society for the Promotion of Science (JSPS),\\
German Israeli Bi-national Science Foundation (GIF), \\
Bundesministerium f\"ur Bildung und Forschung, Germany, \\
National Research Council of Canada, \\
Research Corporation, USA,\\
Hungarian Foundation for Scientific Research, OTKA T-029328, 
T023793 and OTKA F-023259.\\


\begin{thebibliography}{99}

\bibitem{pdg98}
Particle Data Group, C.~Caso~\etal, \EPJ{3}{1998}{1}.

\bibitem{btheo}
See for example: I.I.~Bigi and P.J.~Dornan, \PRP{289}{1997}{1} and 
references therein.

\bibitem{vcbrev}
See for example: K.~\"Osterberg, `Measurement of \mvcb\ and \mvub\ at LEP',
to appear in proceedings of the International
Europhysics Conference on High Energy Physics, Tampere, Finland, 15-21 July 
1999, published by IOP (Bristol, UK).

\bibitem{opalbpbz}
OPAL collaboration, G. Abbiendi~\etal, \EPJ{12}{2000}{609}.

\bibitem{topblife}
DELPHI collaboration, W.~Adam~\etal, \ZPC{68}{1995}{363};\\
L3 collaboration, M.~Acciarri~\etal, \PLB{438}{1998}{417};\\
SLD collaboration, K.~Abe~\etal, \PRL{79}{1997}{590}.

\bibitem{delb0life}
DELPHI collaboration, P.~Abreu~\etal, \ZPC{74}{1997}{19}.

\bibitem{delvcb}
DELPHI collaboration, P.~Abreu~\etal, \ZPC{71}{1996}{539}.

\bibitem{opalvcb}
OPAL collaboration, G.~Abbiendi~\etal, 
`Measurement of \mvcb\ using \bztodslv\ decays',
CERN-EP-2000-032, accepted by Phys.\ Lett.\ B.

\bibitem{deldmd}
DELPHI collaboration, P.~Abreu~\etal, \ZPC{76}{1997}{579}.

\bibitem{dmdth}
See for example: M.~Neubert, \IJA{11}{1996}{4173}.

\bibitem{opaldet}
OPAL collaboration, K.~Ahmet~\etal, \NIM{A305}{1991}{275};\\
P.P.~Allport~\etal, \NIM{A324}{1993}{34};\\
P.P.~Allport~\etal, \NIM{A346}{1994}{476};\\
S.~Anderson~\etal, \NIM{403}{1998}{326}.

\bibitem{jetset}
T.~Sj\"ostrand, \CPC{82}{1994}{74}.

\bibitem{opalrb}
OPAL collaboration, G.~Abbiendi \etal, \EPJ{8}{1999}{217}.

\bibitem{jetcone}
OPAL collaboration, R.~Akers~\etal, \ZPC{63}{1994}{197}.

\bibitem{muonid}
OPAL collaboration, P.D.~Acton~\etal, \ZPC{58}{1993}{523}.

\bibitem{beamspot}
OPAL collaboration, P.D.~Acton~\etal, \ZPC{59}{1993}{183};\\
OPAL collaboration, R.~Akers~\etal, \PLB{338}{1994}{497}.

\bibitem{opaldil} 
OPAL collaboration, R.~Akers~\etal, \ZPC{66}{1995}{555}. 

\bibitem{opalmt}
OPAL collaboration, K.~Ackerstaff~\etal, \EPJ{2}{1998}{213};\\
OPAL collaboration, G.~Abbiendi~\etal, \EPJ{12}{1999}{567}.

\bibitem{opalbinc}
OPAL collaboration, K.~Ackerstaff~\etal, \ZPC{73}{1997}{397}.

\bibitem{opaljpks}
OPAL collaboration, K.~Ackerstaff~\etal, \EPJ{5}{1998}{379}.

\bibitem{opaljetc}
OPAL collaboration, R.~Akers~\etal, \PLB{327}{1994}{411}.

\bibitem{pdg99}
Particle Data Group, 
1999 off-year partial update for the 2000 edition available on 
the PDG WWW pages at {\tt http://pdg.lbl.gov/}.

\bibitem{alephdss}
ALEPH collaboration, D.~Buskulic~\etal, \ZPC{73}{1997}{601}.

\bibitem{opaldprod}
OPAL collaboration, K.~Ackerstaff~\etal, \EPJ{1}{1998}{439}.

\bibitem{ligeti}
A.~Leibovich, Z.~Ligeti, I.~Stewart and M.~Wise, \PRD{57}{1998}{308}.

\bibitem{hfew}
The LEP collaborations, ALEPH, DELPHI, L3 and OPAL,
\NIM{A378}{1996}{101}.\\
Updated averages are described in `Presentation of LEP Electroweak
Heavy Flavour Results for Summer 1998 Conferences', LEPHF 98-01
(see {\tt http://www.cern.ch/LEPEWWG/heavy/} ).

\bibitem{sldbfrag}
SLD collaboration, K.~Abe~\etal, \PRL{84}{2000}{4300}.

\bibitem{fragall}
C.~Peterson, D.~Schlatter, I.~Schmitt and P.~Zerwas, \PRD{27}{1983}{105};\\
P.~Collins and T.~Spiller, \JPH{G11}{1985}{1289};\\
V.G.~Kartvelishvili, A.K.~Likhoded and V.A.~Petrov, \PLB{78}{1978}{615};\\
B.~Anderson, G.~Gustafson and B.~S\"oderberg, \ZPC{20}{1983}{317}.

\bibitem{markIII}
MARK III collaboration, D.~Coffman~\etal, \PLB{263}{1991}{135}.

\bibitem{opallqjet}
OPAL collaboration, K.~Ackerstaff~\etal, \ZPC{76}{1997}{401}.

\bibitem{opaltbex}
OPAL collaboration, R.~Akers~\etal, \ZPC{67}{1995}{379}.

\bibitem{othertb0}
ALEPH collaboration, D.~Buskulic~\etal, \ZPC{71}{1996}{31};\\
CDF collaboration, F.~Abe~\etal, \PRD{58}{1998}{092002};\\
CDF collaboration, F.~Abe~\etal, \PRD{57}{1998}{5382};\\
DELPHI collaboration, P.~Abreu~\etal, \ZPC{68}{1995}{13}.

\bibitem{opalexdmd}
OPAL collaboration, G.~Alexander~\etal, \ZPC{72}{1996}{377}.

\bibitem{otherdmd}
ALEPH collaboration, D.~Buskulic~\etal, \ZPC{75}{1997}{397};\\
CDF collaboration, F.~Abe~\etal, \PRL{80}{1998}{2057};\\
CDF collaboration, F.~Abe~\etal, \PRD{59}{1999}{032001};\\
CDF collaboration, F.~Abe~\etal, \PRD{60}{1999}{072003};\\
CDF collaboration, F.~Abe~\etal, \PRD{60}{1999}{051101};\\
CDF collaboration, F.~Abe~\etal, \PRD{60}{1999}{112004};\\
L3 collaboration, M.Acciarri~\etal, \EPJ{5}{1998}{195}.

%\bibitem{comb}
%ALEPH, CDF, DELPHI, L3, OPAL and SLD collaborations, `Combined results
%on b hadron production rates, lifetimes, oscillations and semileptonic
%decays', CERN-EP-2000-xxx.

\end{thebibliography}
\end{document}